\documentclass[11pt,catalan]{article}
\usepackage{graphicx}
\usepackage{amsmath}
\pdfoutput=1
\usepackage{amsfonts}
\usepackage{amssymb}
\usepackage{bm}

\usepackage[english]{babel}
\usepackage[latin1]{inputenc}
\usepackage{times}
\usepackage[T1]{fontenc}

\newcommand{\RR}{\ensuremath{ \mathbb{R}} }
\newcommand{\NN}{\ensuremath{ \mathbb{N}} }

\newcommand{\CC}{\ensuremath{ \mathbb{C}} }

\newcommand{\ri}{{\rm i} }

\newcommand{\sign}{{\rm sign}}
\newcommand{\tridq}{\stackrel{\ldots}{q}}
\newcommand{\D}{{\rm d}}

\newcommand{\imap}{{\mathbb{I}}{\rm m}}
\newcommand{\realp}{{\mathbb{R}}{\rm e}}

\newtheorem{proposition}{Proposition}

\begin{document}
\title{Non-local Lagrangian Mechanics: Noether theorem and Hamiltonian formalism}
\author{Carlos Heredia\thanks{e-mail address: carlosherediapimienta@gmail.com}\,\,\, and Josep Llosa\thanks{e-mail address: pitu.llosa@ub.edu}\\
Facultat de F\'{\i}sica (FQA and ICC) \\ Universitat de Barcelona, Diagonal 645, 08028 Barcelona,Catalonia, Spain }
\maketitle

\begin{abstract}
We study Lagrangian systems with a finite number of degrees of freedom that are non-local in time. We obtain an extension of Noether theorem and Noether identities to this kind of Lagrangians. A Hamiltonian formalism is then set up for this systems. $n$-order local Lagrangians can be treated as a particular case and the standard results for them are recovered. 
The method is then applied to several other cases, namely two examples of non-local oscillators and the $p$-adic particle. 
\\[1ex]
\noindent
\end{abstract}

\section{Introduction  \label{S0}}
Motivated by improving the ultraviolet (UV) behavior of quantum field theories and solving the cosmological and black holes singularities, non-local models in physics are currently being studied with noteworthy intensity. A brief reading of the literature on this topic already shows the extensive diversity in which are involved: string theory \cite{1}, non-commutative theories \cite{2}, $p$-adic strings \cite{3}, and modify gravity \cite{10}, among others. 

A significant advance concerning the UV problem was the presence of infinite derivatives in the Lagrangian. We observe that, by adding these infinite derivatives to the Lagrangian, the theory's behavior in the UV regime improves \cite{5} without introducing new degrees of freedom to the system  \cite{6}. Furthermore, thanks to this approach, it was conceivable to avoid the Ostrogradsky instabilities \cite{Ostrogradski} that arise when we try to build a Hamiltonian formalism for Lagrangians with a finite number of derivatives. 

%%%%%%%%%%%%%%%%%%%%%%%%%%%%%%%%%
This paper aims to set up a Hamiltonian formalism for non-local Lagrangians extending and improving previous results \cite{Llosa1994}. This new approach considers Lagrangians that explicitly depend on time, which was lacking in the previous approach, as Ferialdi et al. \cite{Ferialdi2012} pointed out.

In the standard case of a first-order local Lagrangian, the Legendre transformation is defined by the momenta. They arise as boundary terms in the integration by parts that one performs when exploiting the variational principle. In a non-local theory and an infinite-order one, it is unclear what those boundary terms are: as there is no highest order derivative of the coordinates, the integration by parts process makes no sense. Nevertheless, in the standard local case, the boundary terms also have a prominent role in the conserved quantities predicted by Noether's theorem when the Lagrangian is invariant by a continuous group of transformations. In our approach for non-local Lagrangians, we shall first work on an extension of Noether's theorem \cite{Noether} and then use the conserved quantity to infer a suitable definition for the momenta.

We shall first establish the variational principle for a non-local Lagrangian, which may depend explicitly on time, and derive the Lagrange equations. As a rule, the latter are of integro-differential type --- or difference-differential in the best case. As we do not have general theorems of existence and uniqueness for these kinds of equations, it makes no sense to take them as laws ruling the system's time evolution from a set of initial data. Instead, we will take them as constraints selecting the class of dynamical trajectories as a submanifold of the broader class of kinematical trajectories. We will especially emphasize what is meant by the time evolution of a given trajectory and we will relate it with the standard notion of evolution in local mechanics.

In Section \ref{S2}, we prove an extension for non-local Lagrangians of Noether's theorem and Noether's identities that follow from the invariance of the Lagrangian under infinitesimal transformations. In Section \ref{S4}, we set up a Hamiltonian formalism for a non-local Lagrangian. We cannot proceed like in the standard method for the local case where the Legendre transformation is a change of coordinates  ---replacing velocities with momenta--- in the space of initial data. Furthermore, due to the lack of theorems of existence and uniqueness for the non-local Lagrange equations, we neither know what the dynamic space is like in the non-local case nor have a system of coordinates for it.

Instead, we will introduce a trivial Hamiltonian formalism on the cotangent space $T(\mathcal{K})$ on the infinite-dimensional manifold of kinematic trajectories and then translate it into a Hamiltonian formalism on the space $\mathcal{D}$ of dynamic trajectories. The instrument will be the pullback of the injection mapping $\mathcal{D}$ into $T(\mathcal{K})$. 

There are two ways of translating the Hamiltonian formalism from a larger phase space to a submanifold. One is based on the Dirac theory of constraints \cite{Dirac1964}, \cite{Sudarshan}. This method implies computing the Poisson brackets between constraints, identifying the first-class constraints, inverting the matrix of second class constraints, and taking the associated Dirac brackets as the effective Poisson brackets for functions defined on the reduced space. Although the whole thing is conceptually simple for a finite-dimensional phase space and a finite number of constraints, it is not so when both the dimensions and the constraints are infinite in number. The other method is based on symplectic mechanics  \cite{Marsden}, \cite{Choquet} --- the covariant counterpart of the Poisson bracket. The Hamiltonian and the symplectic form on the reduced space $\mathcal{D}$ are derived by pulling back the Hamiltonian and the symplectic form on the larger space $T(\mathcal{K})$. This second procedure has the advantage of reaching further with no need to find a system of coordinates for the dynamic space $\mathcal{D}$, which is the most challenging part of the problem since it implies finding a complete system of parameters to characterize the class of all solutions of an integro-differential system of equations. Finding this coordinate system, we would allow us to prove a kind of theorem of existence and uniqueness for such a system.

We start Section \ref{S5} by setting forth a sort of "user's manual" so that the method devised in Section \ref{S4} can be applied as a routine to each individual Lagrangian, with no need for any additional brilliant idea or interpretation. Then we move on to present a list of applications, namely (a) a local Lagrangian of $n$-th order as a particular case, (b) two kinds of non-local oscillators whose dynamic space is either finite-dimensional or infinite-dimensional and (c) a non-local Lagrangian that is somehow related to the $p$-adic string \cite{Sen2000}, \cite{Moeller2002}.

\section{Non-local Lagrangian mechanics \label{S1}}
Consider a dynamic system ruled by the non-local action 
$$  S = \int_{\mathbb{R}} \mathcal{L}\left([q^\alpha],t\right)\,\D t $$
which we will refer to as non-local because the Lagrangian $\mathcal{L}$ may depend on all the values $q^\alpha(\tau)\,, \;\; \alpha=1\ldots m$, at times $\tau$ other than $t$. 

The class of all (possible) kinematic trajectories is the function space $\mathcal{K}= \mathcal{C}^\infty(\mathbb{R},\mathbb{R}^m)$, which we shall call {\em kinematic space}. For time-dependent Lagrangians, we must resort to the {\em extended kinematic space}, $\mathcal{K}^\prime= \mathcal{K} \times\mathcal{R}$. The (non-local) Lagrangian is a real-valued function defined on this space: 
$$ (q^\alpha,t) \in \mathcal{K}^\prime \longrightarrow \mathcal{L}(q^\alpha,t) \in \RR \,.$$
It may depend on all the values $q^\alpha(\tau)\,, \tau\in\RR\,$, not only on $q^\alpha(t)$ and a finite number of derivatives at $t$; however, to make the notation lighter, we shall write $\mathcal{L}(q^\alpha,t)$ instead of $\mathcal{L}([q^\alpha],t)$ as it is usual in most textbooks. With the same purpose, we shall write $q$ instead of $q^\alpha$ provided that there is no risk of misunderstanding. 

The function $q(\tau)$ contains all necessary information about time evolution in the space $\mathcal{K}^\prime$, namely
\begin{equation}  \label{L1} 
  (q,t) \stackrel{T_\tau}{\longrightarrow} (T_\tau q, t+\tau) \,,\qquad {\rm where} \qquad T_\tau q(\sigma) = q(\tau+\sigma)
\end{equation}
It is worth noticing the additive property $\; T_{\tau_1}\circ T_{\tau_2} = T_{\tau_1+\tau_2}\,$.

The action integral is currently understood as 
\begin{equation}   \label{A1}
 S(q) := \int_\RR \D \tau\, \mathcal{L}\left(T_\tau q, \tau \right) \,.
\end{equation}
It may be divergent because the integration spans over an unbounded domain, but we assume that the variation \cite{Gelfand}
$$  \delta S(q) = \int_\RR \D \tau\, \int_\RR \D \sigma\,\frac{\delta \mathcal{L}\left(T_\tau q, \tau \right)}{\delta q(\sigma)} \,{\delta q(\sigma)} $$
is summable for all $\delta q(\sigma)$ with compact support. Then the Lagrange equation is
\begin{equation}  \label{L2o} 
\psi(q,\sigma) = 0 \,, \quad {\rm with} \quad \psi(q,\sigma) := \int_\RR \D \tau\,\lambda(q,\tau,\sigma) \quad {\rm and} \quad
\lambda(q,\tau,\sigma) := \frac{\delta \mathcal{L}\left(T_\tau q, \tau \right)}{\delta q(\sigma)} 
\end{equation}
We call {\em dynamic trajectories} those fulfilling the latter equation.

This presentation of the variational principle is somehow particular in that it is limited to trajectories $(T_\tau q, \tau)\in \mathcal{K}^\prime\,$  corresponding to initial points of the kind $(q,0)$. Nevertheless, a modification of the variational principle based on the trajectory starting at $(q,t)$, for any value of $t$, is more convenient for Lagrangians that explicitly depend on time. Notice that the trajectory starting in $(q, t)$ coincides with the one starting at $(T_{-t}q,0)$ but advanced in $t$, namely
\begin{equation} \label{L1a}
 (T_\tau q, t+\tau) =  (T_{\tau^\prime} \tilde{q}, \tau^\prime)\,, \qquad {\rm with}
\qquad \tau^\prime = t+\tau \quad {\rm and} \quad  \tilde{q}= T_{-t} q 
\end{equation}
Hence the dynamic trajectory starting at $(q,t)$ fulfills
\begin{equation}  \label{L2} 
\Psi(q,t,\sigma) = 0 \,, \quad {\rm where} \quad \Psi(q,t,\sigma) := \psi(T_{-t}q,\sigma+t) 
\end{equation}
or
\begin{equation}  \label{L2z} 
\Psi(q,t,\sigma) := \int_\RR \D \tau\,\Lambda(q,t,\tau,\sigma) \quad {\rm with} \quad
\Lambda(q,t,\tau,\sigma) := \lambda(T_{-t}q,\tau+t,\sigma+t) = \frac{\delta \mathcal{L}(T_\tau q,t+\tau)}{\delta q(\sigma)}
\end{equation}
Later on, we will come back to the meaning of these equations.

Let us see how a standard Lagrangian $L(q,\dot{q},t)$ fits in the theory developed so far. The standard action integral is $\;\;\int \D t\,L(q(t),\dot{q}(t),t)\;$, which has the form (\ref{A1}) provided that we take
$$ \mathcal{L}(T_\tau q,\tau) := L(q(\tau),\dot{q}(\tau),\tau) \,,$$
whence it follows that
$$ \lambda(q,\tau,\sigma) = \left(\frac{\partial L}{\partial q}\right)_{(q,\tau)}\,\delta(\sigma-\tau) - \left(\frac{\partial L}{\partial \dot{q}}\right)_{(q,\tau)}\,\dot\delta(\sigma-\tau) \,, $$
where we have included that $\;q(\tau) = \int_\RR \D\sigma\,q(\sigma)\,\delta(\sigma-\tau)\,$ and $\;\dot{q}(\tau) = -\int_\RR \D\sigma\,q(\sigma)\,\dot\delta(\sigma-\tau)\,$, and have written
$\;\displaystyle{\left(\frac{\partial L}{\partial q}\right)_{(q,\tau)} := \frac{\partial L(q(\tau),\dot{q}(\tau),\tau)}{\partial q}}$, and so on. 
Substituting this in equation (\ref{L2z}), we obtain
\begin{equation}  \label{L2a} 
\Lambda(q,t,\tau,\sigma) = \left(\frac{\partial L}{\partial q}\right)_{(q,t,\tau)}\,\delta(\sigma-\tau) - \left(\frac{\partial L}{\partial \dot{q}}\right)_{(q,t,\tau)}\,\dot\delta(\sigma-\tau) \,, 
\end{equation}
where we have written $\;\displaystyle{\left(\frac{\partial L}{\partial q}\right)_{(q,t,\tau)} = \frac{\partial L(q(\tau),\dot{q}(\tau),t+\tau)}{\partial q}}$, and so on. Substituting (\ref{L2a}) in (\ref{L2z}), we finally arrive at
\begin{equation}  \label{L2b} 
 \Psi(q,t,\sigma) \equiv \frac{\partial L\left(q(\sigma),\dot{q}(\sigma),t+\sigma\right)}{\partial q} -\frac{\D\;}{\D \sigma} \left(\frac{\partial L\left(q(\sigma),\dot{q}(\sigma),t+\sigma\right)}{\partial \dot{q}}\right)
\end{equation}

We must now prove that this is indeed the Lagrange equation for the local Lagrangian $L(q_0,\dot q_0,t)$, namely
\begin{equation}  \label{L2c} 
 U(q_0,\dot q_0,\ddot q_0,t) \equiv \frac{\partial L\left(q_0,\dot q_0,t\right)}{\partial q_0} -\frac{\D\;}{\D \sigma} \left(\frac{\partial L\left(q_0,\dot q_0,t\right)}{\partial \dot{q_0}}\right) = 0
\end{equation}
where we have used $q_0$ for the coordinate value to avoid confusion with $q$ which is  reserved for the whole trajectory. A function $\varphi(\tau)$ is a solution if, and only if,
$$ U(\varphi(\tau),\dot\varphi(\tau),\ddot\varphi(\tau),\tau)  = 0 $$
For a given $t$, we define $\;\sigma = \tau - t\,$ and $q(\sigma) = \varphi(\tau+\sigma)$, and the above equation becomes
$$ U(q(\sigma),\dot q(\sigma),\ddot q(\sigma),t+\sigma)  = 0 $$
which, including the definition in (\ref{L2c}), is the equation (\ref{L2b}). \hfill $\Box$

%%%%%%%%%%%%%%%%%%%%%%%%%%%%%%%%%%%%%%%%%%%%%%%%%%%%%%%%%%%
\subsection{Lagrange equations and time evolution \label{S1.1}}
Of course, equations (\ref{L2}) are not fulfilled by all kinematic trajectories $q\in\mathcal{K}$. Therefore, they can be taken as implicit equations by defining the {\em extended dynamic space} $\mathcal{D}^\prime$, i. e. the class of all dynamic trajectories, as a submanifold of $\mathcal{K}^\prime = \mathcal{K}\times \RR$.

In the local first-order case, Lagrange equations (\ref{L2}) turn out to be an ordinary differential system of the second order (\ref{L2b}) that can be solved in the accelerations $\ddot{q} $ as functions of coordinates, velocities, and time\footnote{Similarly, the Lagrange equations for a regular local order $n$ Lagrangian are an ordinary differential system of order $2 n$. }. The theorems of existence and uniqueness then imply that, given the coordinates and velocities at an initial time  $(q_0,\dot{q}_0, t_0)$, there is a unique solution
\begin{equation} \label{A20}
 q(\sigma) = \varphi(q_0,\dot{q}_0,t_0;\sigma) \qquad\mbox{such that} \qquad 
q_0 = \varphi(q_0,\dot{q}_0,t_0;0)\,,\qquad  \dot{q}_0 = \partial_\sigma\varphi(q_0,\dot{q}_0,t_0;0) 
\end{equation}
This fact is usually read as though Lagrange equations rule the evolution of the system. Consequently, every trajectory in the extended dynamic space $\mathcal{D}^\prime$ may be labeled with those $2 n + 1$ coordinates, and it is identified with the {\em initial data space}.

The case of a non-local Lagrangian is not so simple \cite{Barnaby2008}, \cite{Calcagni} because equations (\ref{L2}) are usually of integro-differential type and, as a rule, there are no theorems of existence and uniqueness supporting the above interpretation in terms of evolution from a set of initial data. Moreover, the extended dynamic space $\mathcal{D}^\prime$ may have an infinite number of dimensions.

The latter leads us to propose an alternative view and take the Lagrange equations (\ref{L2}) as the constraints that define $\mathcal{D}^\prime$ as a submanifold of $\mathcal{K}^\prime$ in implicit form. This picture holds for the standard local case as well. However, the theorems of existence and uniqueness imply that the shape of the dynamic trajectories is (\ref{A20}), namely the explicit parametric equations of the submanifold $\mathcal{D}^\prime$. Thus these theorems determine the number of essential parameters to individualize a dynamic solution, i. e. to coordinatize the dynamic space.
 
We shall write the constraints as
\begin{equation}  \label{L3} 
 \Psi\left(q, t\right) = 0 \,, \qquad \qquad \Psi\left(q, t\right)_{(\sigma)} := 
\int_\RR \D \tau\,\Lambda(q,t,\tau,\sigma)
\end{equation}
The notation is meant to suggest that $\Psi$ maps the extended kinematic space onto a space of smooth functions of the real variable $\sigma$ and the trajectories in $\mathcal{D}^\prime$ are those $(q,t)$ such that make $\psi $ null.

The generator of time evolution (\ref{L1}) in $\mathcal{K}^\prime$ is the field of vectors that are tangent to the curves $(T_\tau q, t+\tau)\,$, and thus, for a function $F(q,t)$, we have that
\begin{equation}  \label{L4} 
\mathbf{X} F(q,t) := \left[\frac{\partial F\left(T_\varepsilon q, t+\varepsilon \right)}{\partial \varepsilon} \right]_{\varepsilon=0}
\end{equation}
Moreover, we can write the generator as
\begin{equation}  \label{L5} 
\mathbf{X} := \partial_t  + \int \D\sigma\,\dot{q}(\sigma)\,\frac{\delta \quad }{\delta q(\sigma)} 
\end{equation}
where $\dot{q}$ means the derivative function, by the chain rule.

By its very construction, the constraints (\ref{L3}) are stable under time evolution or, what is equivalent, the vector $\mathbf{X}$ must be tangent to the dynamic space $\mathcal{D}^\prime$. Indeed from (\ref{L3}) and (\ref{L2}), we have that 
$$  \Psi\left(T_\varepsilon q, t+\varepsilon \right)_{(\sigma)} = \int \D\tau\,\frac{\delta  L\left(T_{\tau+\varepsilon} q, t+\varepsilon+\tau \right)}{\delta T_\varepsilon q(\sigma)} = \int \D\tau^\prime\,\frac{\delta  L\left(T_{\tau^\prime} q, t+\tau^\prime \right)}{\delta q(\sigma+\varepsilon)} = \Psi\left( q, t \right)_{(\sigma+\varepsilon)} \,,$$
where  $\;\tau^\prime=\tau+\varepsilon\,$.
Thus if $\Psi(q,t)_{(\sigma)}=0\,$ for all $\sigma$, then $\Psi\left(T_\varepsilon q, t+\varepsilon \right)_{(\sigma)} =0$ as well and therefore
$$ \mathbf{X} \Psi(q,t)_{(\sigma)} = \left[\frac{\partial \Psi\left(T_\varepsilon q, t+\varepsilon \right)_{(\sigma)}}{\partial \varepsilon} \right]_{\varepsilon=0} = 0 $$

\section{Noether's theorem  \label{S2}}
The proof of Noether's theorem for a local Lagrangian containing derivatives up to the $n$-th order involves some integrations by parts to remove the derivatives of $\delta q(t)$ of orders higher than $n-1$. As a consequence, the variation of the action contains some boundary terms that, in the end, give rise to a conserved quantity. In a non-local Lagrangian, there is no such highest order derivative, and the latter scheme does not make sense. We shall use a trick we previously used in \cite{Heredia} in the context of Classical Field Theory that will bring out the equivalent of those boundary terms without resorting to integrations by parts.

Consider an infinitesimal transformation
\begin{equation}  \label{A3}
t^\prime(t) = t + \delta t(t) \,, \qquad \quad q^{\prime}(t) = q(t) + \delta q(t) \,.
\end{equation}
and let $[t_1,t_2]$ be a time interval and $[t^\prime_1,t^\prime_2]$ the transformed interval according to (\ref{A3}). We define 
\begin{equation}  \label{A3b}
 \Delta S(q;t_1,t_2) := \int_{t^\prime_1}^{t^\prime_2} \mathcal{L}\left(T_{t^\prime}q^\prime,t^\prime\right)\,\D t^\prime - \int_{t_1}^{t_2} \mathcal{L}\left(T_t q,t\right)\,\D t 
\end{equation}
If the transformation (\ref{A3}) leaves invariant the action, then $\Delta S(q,t_1,t_2)$ vanishes.

Replacing with $\tau$ the dummy variables $t$ and $t^\prime$ in the integrals on the right-hand side, we have that
\begin{eqnarray*}  
 \Delta S(q;t_1,t_2) &=& \int_{t^\prime_1}^{t^\prime_2} \mathcal{L}\left(T_\tau q^\prime,\tau\right)\,\D \tau -\int_{t_1}^{t_2} \mathcal{L}\left(T_\tau q,\tau\right)\,\D \tau 
 \\[1ex]   
  &=& \mathcal{L}\left(T_{t_2}q,t_2\right)\,\delta t_2 - \mathcal{L}\left(T_{t_1}q,t_1\right)\,\delta t_1 + \int_{t_1}^{t_2} \D \tau \, \left[\mathcal{L}\left(T_\tau q^\prime ,\tau\right) - \mathcal{L}\left(T_\tau q,\tau\right) \right] 
 \\[1ex]   
  &=&  \int_{t_1}^{t_2} \D \tau \,\left[\frac{\partial\quad }{\partial \tau} \left[\mathcal{L}\left(T_{\tau}q,\tau\right)\,\delta t(\tau)\right]+ \int_\RR \D\sigma\,\frac{\delta \mathcal{L}\left(T_\tau q, \tau \right)}{\delta q(\sigma)}\,\delta q(\sigma) \right]\,,
\end{eqnarray*}
which, adding and subtracting (\ref{L2o}) to the right-hand side, becomes
\begin{eqnarray*}
\lefteqn{\Delta S(q;t_1,t_2) - \int_{t_1}^{t_2} \D \tau \,\psi(q,\tau)\,\delta q(\tau) = \int_{t_1}^{t_2} \D \tau \,\left[\frac{\partial\quad }{\partial \tau} \left[\mathcal{L}\left(T_{\tau}q,\tau\right) \delta t(\tau)\right]\right. }\\[1ex]
  & & \hspace*{6em} \left. +\int_\RR \D\sigma\,\lambda(q,\tau,\sigma)\,\delta q(\sigma) - \int_\RR \D\sigma\,\lambda(q,\sigma,\tau)\,\delta q(\tau) \right]  
\end{eqnarray*}
and, after suitable changes of the variable $\sigma$ in the last two integrals, we arrive at
\begin{eqnarray}  
 \Delta S(q;t_1,t_2) &=& \int_{t_1}^{t_2} \D \tau \,\left\{\psi(q,\tau)\,\delta q(\tau) +\frac{\partial \;}{\partial \tau} \left[\mathcal{L}\left(T_{\tau}q,\tau\right)\,\delta t(\tau)\right]+ \right. \nonumber \\[2ex]  \label{A4}
 & & \left.\qquad \int_\RR \D\xi\,\left[\lambda(q,\tau,\tau+\xi)\,\delta q(\tau+\xi) - \lambda(q,\tau-\xi,\tau)\,\delta q(\tau) \right]\right\} 
\end{eqnarray}
Now we can write the integrand in the last term on the right-hand side as
\begin{eqnarray*}
\lefteqn{\lambda(q,\tau,\tau+\xi)\,\delta q(\tau+\xi) -\lambda(q,\tau-\xi,\tau)\,\delta q(\tau) =} \\[1ex]
 \qquad &  \qquad \quad= &  \int_0^1\D\eta \,\frac{\partial\;}{\partial\eta}\left[\lambda(q,\tau+(\eta-1)\xi,\tau+\eta\xi)\,\delta q(\tau+\eta\xi) \right] \\[2ex]
 &  \qquad \quad = & \xi\, \frac{\partial\;}{\partial \tau}\,\int_0^1\lambda(q,\tau+(\eta-1)\xi,\tau+\eta\xi)\,\delta q(\tau+\eta\xi)\,\D\eta 
\end{eqnarray*}
which substituted in (\ref{A4}) yields
\begin{equation}  \label{A5}
 \Delta S(q;t_1,t_2) = \int_{t_1}^{t_2} \D \tau \,\left[\psi(q,\tau)\,\delta q(\tau) +\frac{\partial }{\partial \tau} \left\{\mathcal{L}\left(T_{\tau}q,\tau\right)\,\delta t(\tau) + \Pi(q,\tau) \right\}\right] 
\end{equation}
with
\begin{equation}   \label{A6}
\Pi(q,\tau) = \int_\RR\D\xi\,\xi\,\int_0^1\,\lambda(q,\tau+(\eta-1)\xi,\tau+\eta\xi)\,\delta q(\tau+\eta\xi) \,\D\eta 
\end{equation}
Replacing now $\tau+\eta\xi=\rho$,  the latter can be written as
$$ \Pi(q,\tau) = \int_\RR \D\xi\,\int_\tau^{\tau+\xi}\lambda(q,\rho-\xi,\rho)\,\delta q(\rho)\,\D\rho  $$
which, after inverting the order of the integrals, leads to 
$$ \Pi(q,\tau) = \int_\RR \D\rho\,\delta q(\rho)\, \int_\RR\D\xi\,\left[\theta(\rho-\tau)\,\theta(\tau+\xi-\rho)- \theta(\tau-\rho)\,\theta(\rho-\tau-\xi)\right]\,\lambda(q,\rho-\xi,\rho) $$
or
$$ \Pi(q,\tau) = \int_\RR \D\rho\,\delta q(\rho)\, \int_\RR\D\xi\,\left[\theta(\tau+\xi-\rho)-\theta(\tau-\rho)\right]\,\lambda(q,\rho-\xi,\rho) $$
where $\theta$ is the Heaviside step function and, on replacing $\rho-\xi = \zeta$, it becomes
\begin{equation}   \label{A7}
\Pi(q,\tau) = \int_\RR \D\rho\,\delta q(\rho)\, P(q,\tau,\rho) \,,\quad {\rm with} \quad P(q,\tau,\rho):= \int_\RR\D\zeta\,\left[\theta(\tau-\zeta)- \theta(\tau-\rho)\right]\,\lambda(q,\zeta,\rho)
\end{equation}
The resemblance of $\Pi(q,\tau)$ with the ``boundary terms'' that one encounters in Noether's theorem for standard Lagrangians is obvious.

Finally, substituting this in equation (\ref{A5}), we arrive at $\;
 \Delta S(q;t_1,t_2) = \int_{t_1}^{t_2} \D \tau \,N(q,\tau) \,$,
with
\begin{equation}  \label{A8}
 N(q,\tau):=\psi(q,\tau)\,\delta q(\tau) +\frac{\partial\; }{\partial \tau}\left[\mathcal{L}\left(T_{\tau}q,\tau\right)\,\delta t(\tau)+ \int_\RR \D\rho\,\delta q(\rho)\, P(q,\tau,\rho) \right] 
\end{equation}

Provided that the Lagrangian is invariant under the transformation (\ref{A3}), the definition (\ref{A3b}) implies that $\,\Delta S(q;t_1,t_2) = 0\,$, for any $t_1,\, t_2\,$, and therefore equation (\ref{A8}) implies that
\begin{equation}  \label{A9}
 N(q,\tau):=\psi(q,\tau)\,\delta q(\tau) +\frac{\partial\; }{\partial \tau} \left[\mathcal{L}\left(T_{\tau}q,\tau\right)\,\delta t(\tau)+\int_\RR \D\rho\,\delta q(\rho)\,P(q,\tau,\rho) \right] \equiv 0 
\end{equation}
which is an extension of the {\em Noether's identity} to non-local Lagrangians. It is said to be an {\em off-shell} equality because it holds for all trajectories regardless of whether they are dynamic or not. 

The latter identity has been obtained for those trajectories starting at $(q,0)\in\mathcal{K}^\prime\,$. In order to extend it to trajectories that start at any $(q,t)\in\mathcal{K}^\prime$, we invoke (\ref{L1a}) and define
$$\; N(q,t,\tau):= N\left(T_{-t}q,t+\tau\right) \,.$$ 
Then, including (\ref{L2}), (\ref{L2z}) and (\ref{A11}), the identity (\ref{A9}) becomes
\begin{eqnarray}  \label{A10}
N(q,t,\tau) &:=& \Psi(q,t,\tau)\,\delta q(\tau) +\frac{\partial J(q,t,\tau)}{\partial \tau}  \equiv 0 \,, \\[1ex]  \label{A10a}
{\rm where} \quad\qquad J(q,t,\tau) &:=& \mathcal{L}\left(T_{\tau}q,t+\tau\right)\,\delta t(t+\tau)+ \int_\RR \D\rho\,\delta q(\rho)\, P(q,t,\tau,\rho)   \\[1ex]  \label{A11}
{\rm with} \qquad P(q,t,\tau,\rho) &:=& \int\D\zeta\, \left[\theta(\rho-\tau)-\theta(\zeta-\tau)\right]\,\Lambda(q,t,\zeta,\rho)  \,. 
\end{eqnarray}
%with
%\begin{equation}  \label{A11}
%\end{equation}
It can be easily proved from the definition (\ref{L2z}) that $\Lambda(T_\varepsilon q,t+\varepsilon,\zeta,\rho) = \Lambda(q,t,\zeta +\varepsilon,\rho +\varepsilon)$, whence it follows that
$\; J\left(T_\varepsilon q,t+\varepsilon,\tau\right) = J(q,t,\tau+\varepsilon) \,$,
which implies that 
\begin{equation}  \label{A12a}
\mathbf{X} J(q,t,\tau) = \frac{\partial\,}{\partial \tau} J(q,t,\tau)  
\end{equation}
where  $\;\mathbf{X}\;$ is the generator (\ref{L4}) of time evolution in $\mathcal{D}^\prime\,$.

Equation (\ref{A10}) is the general expression of Noether's identity for a non-local Lagrangian that may explicitly depend on $t$. For dynamic trajectories ({\em on-shell}), i. e. solutions of the Lagrange equations (\ref{L2}), this identity implies that
$$ \frac{\partial \;}{\partial \tau} J(q,t,\tau) = 0 \qquad\qquad {\rm and}  \qquad\qquad   
\mathbf{X} J(q,t,\tau) = 0 $$
As a consequence, $ J(q,t,\tau)$ does not depend on $\tau$, and the quantity
\begin{equation}  \label{A12}
J(q,t):= J(q,t,0) = \mathcal{L}\left(q,t\right)\,\delta t(t)+ \int_\RR \D\rho\,\delta q(\rho)\, P(q,t,0,\rho)
\end{equation}
is preserved by time evolution. 

If the Lagrangian does not explicitly depend on $t$, it is invariant by time translations, and the above results can be applied to the infinitesimal transformation
\begin{equation}  \label{A3a}
 \delta(t) = \varepsilon \qquad {\rm and} \qquad \delta q(t) = - \varepsilon \,\dot{q} \,,
\end{equation}
We call the {\em energy} the preserved quantity $\;E = - J(q,t)/\varepsilon\,$, that is
\begin{equation}  \label{A12z}
E(q,t):=  - \mathcal{L}\left(q,t\right) + \int_\RR \D\rho\,\dot q(\rho)\, P(q,t,0,\rho)
\end{equation}

Let us now particularize these results for a standard local first-order Lagrangian $L(q,\dot{q},t)$. Substituting (\ref{L2a}) in the definition (\ref{A11}), we have that
\begin{eqnarray}
  P(q,t,\tau,\rho) & = & \int_\RR \D \zeta\,\left[\theta(\tau-\zeta)-\theta(\tau-\rho)\right]\, \left[ \left(\frac{\partial L}{\partial q}\right)_{(q,t,\zeta)}\hspace*{-.5em}\delta(\rho-\zeta) - \left(\frac{\partial L}{\partial \dot{q}}\right)_{(q,t,\zeta)}\hspace*{-.5em}\dot\delta(\rho-\zeta)\right] \nonumber \\[2ex]  \label{A13a}
	& = & \delta(\tau-\rho)\,\left(\frac{\partial L}{\partial \dot{q}}\right)_{(q,t,\tau)} \
\end{eqnarray}
Substituting the latter in (\ref{A10}-\ref{A10a}), Noether's identity (off-shell) becomes 
\begin{equation} \label{A11a}
 N(q,t,\tau) \equiv  \Psi(q,t,\tau)\,\delta q(\tau) + \frac{\partial J(q,t,\tau)}{\partial \tau} =0
\end{equation}
with
\begin{equation}  \label{A11b}
J(q,t,\tau) :=  L(q(\tau),\dot{q}(\tau),t+\tau)\,\delta t(t+\tau) + \frac{\partial L(q(\tau),\dot{q}(\tau),t+\tau)}{\partial\dot{q}}\,\delta q(\tau) 
\end{equation}

The function $J(q,t,\tau)$ on-shell does not depend on $\tau$ and therefore,
\begin{equation}  \label{A11bb}
J(q,t) :=  L(q_0,\dot{q}_0,t)\,\delta t(t) + \frac{\partial L(q_0,\dot{q}_0,t)}{\partial\dot{q}_0}\,\delta q_0 
\end{equation}
is a constant and only depends on the trajectory $q$ through the initial values $(q_0,\dot{q}_0)$ at $t$. 

In turn, the energy is
\begin{equation}  \label{A11c}
E(q,t):=  - L(q_0,\dot{q}_0,t) + \frac{\partial L(q_0,\dot{q}_0,t)}{\partial\dot{q}_0}\,\dot q_0
\end{equation}

\section{Hamiltonian formalism for a non-local Lagrangian  \label{S4}}
Our next aim is to set up a Hamiltonian formalism for the Lagrange equations (\ref{L3}). The standard procedure for  first-order local Lagrangians consists of introducing the canonical momenta and, by inverting the Legendre transformation, replace one-half of the variables, namely the velocities, with the momenta as coordinates in $\mathcal{D}^\prime$. With this new coordinatization, the dynamic space becomes the {\em phase space}.

This is not feasible in the non-local case because: (a) so far we have not coordinated $\mathcal{D}^\prime$ yet ---as a rule it depends on the type of equations (\ref{L3})--- and (b) the fact that $\mathcal{D}^\prime$ likely has an infinite number of dimensions (and, as everyone knows, half of infinity is infinity). However it is worth noticing that, in the local first-order case, (a) the canonical momentum for the Legendre transformation is the prefactor of $\delta q(t)$ in the conserved quantity (\ref{A11bb}). This will be the basis for the ``oriented guess'' that we shall make in the next Section.

\subsection{Legendre transformation \label{S4.1} }
We first introduce the extended phase space $\,\Gamma^\prime =\mathcal{K}^2\times \mathbb{R}\,$ made of points $(q, \pi,t)$, where $q, \,\pi \in\mathcal{K}\,$ are smooth functions, and the Hamiltonian
\begin{equation}  \label{L6} 
H(q,\pi,t) = \int \D\sigma\,\pi(\sigma)\,\dot{q}(\sigma) -  \mathcal{L}(q,t)
\end{equation}
and the Poisson bracket
$$ \left\{ F, G \right\} = \int \D\sigma\,\left(\frac{\delta F}{\delta q(\sigma)}\,\frac{\delta G}{\delta \pi(\sigma)} - \frac{\delta F}{\delta \pi(\sigma)}\,\frac{\delta G}{\delta q(\sigma)} \right) \,.$$
The Hamilton equations are
\begin{eqnarray}  \label{L7a}
\mathbf{H} {q}(\sigma) & = & \frac{\delta H}{\delta \pi(\sigma)} = \dot{q}(\sigma)  \\[1ex]  \label{L7b}
\mathbf{H} {\pi}(\sigma) & = & -\frac{\delta H}{\delta q(\sigma)} = \dot{\pi}(\sigma) + \frac{\delta \mathcal{L}(q,t)}{\delta q(\sigma)} \,,
\end{eqnarray}
where $\mathbf{H}$ is the infinitesimal generator of the Hamiltonian flow 
$$ \mathbf{H} = \partial_t + \int \D\sigma\,\left( \dot{q}(\sigma)  \,\frac{\delta \quad}{\delta q(\sigma)} + \left[\dot{\pi}(\sigma) + \frac{\delta \mathcal{L}(q,t)}{\delta q(\sigma)} \right] \,\frac{\delta \quad}{\delta \pi(\sigma)} \right) \,.$$
Hamilton equations can be written in a more compact form using the contact differential 2-form
\begin{equation}  \label{L7c} 
 \Omega^\prime = \Omega - \delta H \wedge \delta t\,, \qquad {\rm where} \qquad \Omega = \int \D\sigma\;\delta \pi(\sigma) \wedge \delta q(\sigma) \,,
\end{equation}
is the symplectic form \cite{Marsden} (and we have written``$\delta$'' to distinguish the differential on the manifold $\Gamma^\prime$ from the ``$\D$'' occurring in the notation for integrals that we have adopted here). Then Hamilton equations (\ref{L7a}-\ref{L7b}) are equivalent to
\begin{equation}  \label{L7d} 
\ri_{\mathbf{H}} \Omega^\prime = 0
\end{equation}

So far, this Hamiltonian system in the extended phase space $\Gamma^\prime$ has almost nothing to do with the Lagrange equations (\ref{L3}) nor the generator of time evolution $\mathbf{X}$ in the space $\mathcal{D}^\prime\,$. However, we can connect both through the injection 
\begin{equation}  \label{L8} 
 (q, t)\in \mathcal{D}^\prime \stackrel{j}{\longrightarrow} (q,\pi,t)\in\Gamma^\prime \,, \qquad  \mbox{where } \qquad \pi := P(q,t) \in\mathcal{K}  \,,
\end{equation}
and $\, P(q,t)_{(\sigma)} \,$ is the prefactor of $\delta q(\sigma)$ in the Noether conserved quantity (\ref{A12}), that is
\begin{equation}  \label{L8a} 
P(q,t)_{(\sigma)} = P(q,t,0,\sigma) = \int_\RR \D\zeta\,\left[\theta(\sigma)-\theta(\zeta)\right]\,\Lambda(q,t, \zeta,\sigma).
\end{equation}
$j$ defines a 1-to-1 map from $\mathcal{D}^\prime$ into its range, $j(\mathcal{D}^\prime)\subset \Gamma^\prime$, i. e. the submanifold implicitly defined by the constraints
\begin{equation}  \label{L9} 
\Psi\left(q, t\right)= 0  \qquad {\rm and} \qquad 
\Upsilon\left(q, \pi, t\right):= \pi - P(q,t) = 0
\end{equation}

\begin{proposition}     
The Jacobian map $\,j^T\,$ maps the  infinitesimal generator $\mathbf{X}$ of time evolution in $\mathcal{D}^\prime$ into $\mathbf{H}$, the generator of the Hamiltonian flow in $\,\Gamma^\prime\,$. 
\end{proposition}

\paragraph{Proof:} To begin with, we have that
$\;\; \left(j^T\mathbf{X}\right) q(\sigma) = \mathbf{X} q(\sigma) = \dot q(\sigma) = \mathbf{X} q(\sigma)\,,\;$ where (\ref{L9}) and (\ref{L5}) have been included. 
Then 
$$\;\; \left(j^T\mathbf{X}\right) \pi(\sigma) = \mathbf{X} P(q,t,0,\sigma) = \partial_\varepsilon\left[\int_\RR\D\zeta\,\left[\theta(\sigma)-\theta(\zeta)\right]\,\Lambda(T_\varepsilon q,t+\varepsilon,\zeta,\sigma)\right]_{\varepsilon=0}\,.\;$$
Now, including (\ref{L2z}), we have
$$  \Lambda(T_\varepsilon q,t+\varepsilon,\zeta,\sigma) = \frac{\delta \mathcal{L}\left(T_{\zeta+\varepsilon} q, t+\varepsilon+\zeta\right)}{\delta T_\varepsilon q(\sigma)} = \Lambda(q,t,\zeta+ \varepsilon,\sigma+\varepsilon) $$
which, substituted in the integral above and replacing $\,\zeta+ \varepsilon=\zeta^\prime\,$, yields
\begin{eqnarray*} 
\mathbf{X} P(q,t,0,\sigma) &=& \partial_\varepsilon\left[\int_\RR\D\zeta^\prime\, \left[\theta(\sigma)-\theta(\zeta^\prime-\varepsilon)\right]\,\Lambda(q,t,\zeta^\prime,\sigma+\varepsilon)\right]_{\varepsilon=0} \\[1ex]
 & = & \int_\RR\D\zeta\,\left\{-\partial_\zeta \,\left[\theta(\sigma)-\theta(\zeta)\right]\,\Lambda(q,t,\zeta,\sigma) + \left[\theta(\sigma)-\theta(\zeta)\right]\,\partial_\sigma \Lambda(q,t,\zeta,\sigma)\right\}.
\end{eqnarray*}
Furthermore, using that $\;\partial_\zeta \theta(\zeta)=\delta(\zeta) \,$, we arrive at  
$$ \mathbf{X} P(q,t,0,\sigma) = \frac{\delta \mathcal{L}(q,t)}{\delta q(\sigma)} + \partial_\sigma P(q,t,0,\sigma) - \delta(\sigma) \,\Psi(q,t,\sigma) \,,$$
where (\ref{L2z}) has been included. As the point $(q,t)\in\mathcal{D}^\prime$, the last term in the right vanishes by the constraints (\ref{L9}), and we finally obtain
$$ \left(j^T\mathbf{X}\right) \pi(\sigma) = \frac{\delta \mathcal{L}(q,t)}{\delta q(\sigma)} + \partial_\sigma P(q,t,0,\sigma) = \mathbf{H} \pi(\sigma) \hspace*{6em} \Box$$
 
As a corollary $\mathbf{H}= j^T\mathbf{X}$ is tangent to the submanifold $j(\mathcal{D}^\prime)$, and therefore, the constraints (\ref{L9}) are stable by the Hamiltonian flow.

To translate the Hamiltonian formalism in $\Gamma^\prime$ into a Hamiltonian formalism in the extended dynamic space $\mathcal{D}^\prime$ we use that the pullback $j^\ast$ maps the contact form (\ref{L7c}) onto the differential 2-form 
\begin{equation} \label{L10}
\omega^\prime  = J^\ast \Omega^\prime = \int \D\sigma \,\delta P(q,t)(\sigma)\wedge \delta q(\sigma) - \delta h \wedge \delta t \,, \qquad \quad \omega^\prime\in\Lambda^2(\mathcal{D}^\prime)\,,
\end{equation} 
where $\,h = H\circ j\,$. Then as $\,j^T\mathbf{X} = \mathbf{H}\,$, the pullback of equation (\ref{L7d}) implies that  
\begin{equation} \label{L11}
\ri_{\mathbf{X}} \omega^\prime = 0 
\end{equation}

The reduced Hamiltonian $h(q,t)$ and the contact form $\omega^\prime$ on $\mathcal{D}^\prime\,$ can be derived from equations (\ref{L6}), (\ref{L7c}), and (\ref{L8a}), and they are respectively 
\begin{equation}  \label{L12} 
h(q,t) = \int \D\tau\,\D\sigma\,\left[\theta(\sigma)-\theta(\tau)\right]\,\dot q(\sigma)\,\Lambda(q, t,\tau,\sigma) - \mathcal{L}(q,t)\,, 
\end{equation}
and $\;\, \omega^\prime_{(q,t)} = - \delta h(q,t) \wedge \delta t + \omega_{(q,t)} \,$, where
\begin{equation}  \label{L13} 
\omega_{(q,t)} = \frac12\,\int \D\tau\,\D\sigma\,\D\rho\,\frac{\delta \Lambda(q, t,\tau,\sigma)}{\delta q(\rho)} \,\left[\theta(\sigma)-\theta(\rho)\right]\,\delta q(\rho) \wedge \delta q(\sigma) \,,
\end{equation}
is the (pre)symplectic form. To derive the latter expression, we have included the skew-symmetry of $\delta q(\rho) \wedge \delta q(\sigma)$.

We have not reached our goal yet because, due to the constraints $\,\Psi(q,t)=0\,$ that characterize the dynamic space as a submanifold of the kinematic space $\mathcal{K}^\prime\,$, $q$ and $t$ are not independent coordinates in $\mathcal{D}^\prime$\,. Therefore, a final step remains and consists of coordinatizing $\mathcal{D}^\prime$. We need to obtain the explicit parametric form of the submanifold $\mathcal{D}^\prime$ instead of the implicit form provided by the Lagrange equations. This is easy for regular local Lagrangians that depend on derivatives up to the $n$-th order\footnote{Namely, the Hessian determinant concerning highest order derivatives does not vanish.} because Lagrange equations are an ordinary differential system of order $2n$ and the existence and uniqueness theorems provide the sought parametric form (see section \ref{S5.1} below).  
However, in the general case, deriving the explicit equations of $\mathcal{D}^\prime$ from the implicit equations is a complex task that depends on each specific case. This is intended to exemplify in the next Section.

%%%%%%%%%%%%%%%%%%%%%%%%%%%%%%%%%%%%%%%%%%%%%%
\section{Application \label{S5}}
The ``user manual'' for the procedure developed so far would read:
\begin{itemize}
\item To start with, write the action integral so that the function $ \mathcal{L}(T_\tau q,\tau)$ can be identified, 
\item compute the functional derivatives $\,\lambda(q,\tau,\sigma)\,$ and $\Lambda(q,t,\tau,\sigma)$ ---see equations (\ref{L2o}) and (\ref{L2z})---,
\item substitute the latter in (\ref{L13}) and calculate the contact form and 
\item the Hamiltonian (\ref{L12}).
\end{itemize}
In what follows, we shall apply these directions to some specific cases. In each of them, the main difficulty stems from finding a complete set of coordinates to characterize the elements of the dynamic space $\mathcal{D}$. 

%%%%%%%%%%%%%%%%%%%%%%%%%%%%%%%%%%%%%%%%%%%%%%%%%%%%%%%%%%%%%%%%%%%%%%%%%%%%%%%%%%%
\subsection{Regular local Lagrangian of the first order \label{S5.1}}
Let us consider a regular Lagrangian $\mathcal{L}(q,t) = L(q,\dot q,t)\,$, i. e. $\displaystyle{\frac{\partial^2 L}{\partial \dot q^2} \neq 0}\,$. We apply the results in the former Section to determine the canonical momenta and the Hamiltonian in the dynamic space $\mathcal{D}^\prime\,$. 

As seen in Section \ref{S1.1},  $\mathcal{D}^\prime\,$ is coordinated by $(q,p,t)$.
Equation (\ref{L2a}) gives the functional derivative $\;\displaystyle{\frac{\delta \mathcal{L}(T_\tau q,t+\tau) }{\delta q(\sigma)} = \Lambda(q,t,\tau,\sigma) }\;$, which is necessary to calculate formula (\ref{L8a}). It yields
$$\;P(q,t,0,\sigma ) = \delta(\sigma)\,p \qquad {\rm with} \qquad p:=\frac{\partial L(q,\dot q,t)}{\partial \dot q}\,$$
Therefore, the Hamiltonian (\ref{L12}) in the dynamic space $\mathcal{D}^\prime$ is 
$$ h(q,p,t) = p\,\dot q(q,p,t) - L(q,\dot q(q,p,t),t) \,,$$
where $\dot q(q,p,t)$ is the result of solving $\;\displaystyle{ p:=\frac{\partial L(q,\dot q,t)}{\partial \dot q}}\;$, which can be done because the Lagrangian is assumed to be regular. In turn, the contact form (\ref{L10}) is
$$ \omega^\prime = \delta p \wedge \delta q - \delta h\wedge \delta t \in \Lambda^2(\mathcal{D}^\prime)\,.$$
Then the Hamilton equations $\;\ri_\mathbf{X} \omega^\prime = 0\;$are equivalent to
$$  \mathbf{X} q = \dot q(q,p,t)  \qquad {\rm and} \qquad 
\mathbf{X} p = -\frac{\partial h(q,p,t)}{\partial q} = \frac{\partial L(q,\dot q,t)}{\partial q} \,,$$
and they are equivalent to the Lagrange equations $\;\displaystyle{\frac{\partial L(q,\dot q,t)}{\partial q} - \mathbf{X}\left(\frac{\partial L(q,\dot q,t)}{\partial \dot q}\right) = 0 }\,.$

We could proceed similarly with a local regular $n$-th order Lagrangian $L(q,\dot q,\ldots q^{(n)},t)\,$ to obtain
$$ P(q,t)(\sigma) = \sum_{l=0}^{n-1} p_l\,(-1)^{l}\delta^{(l)}(\sigma) \,,$$
where $\delta^{(l)}$ is the $l$-th derivative of Dirac delta function and
 $p_l\,, \quad l=0 \ldots n-1,$ are the Ostrogradski momenta \cite{Whittaker},\cite{Ostrogradski}
$$ p_l = \sum_{k=0}^{n-l-1} (-1)^k\frac{\D^k \;}{\D t^k}\left(\frac{\partial L}{\partial q^{(l+k+1)}}\right) \,,$$
($ q^{(k)}\,$ means the $k$-th derivative).

%%%%%%%%%%%%%%%%%%%%%%%%%%%%%%%%%%%%%%%%%%%%%%%%%%%%%%%%%%%%%%%%%%%%%%%%%%%%%%%%%%%
\subsection{Non-local harmonic oscillator \label{S5.2}}
Consider the action integral 
\begin{equation}  \label{OS1}
S = \int_\RR \D t\,\left[\frac12\,\dot q^2(t) -\frac{\omega^2}2\, q^2(t) + \frac{g} 4 \, q(t)\,\int_\RR \D \zeta \,K(\zeta)\, q(t-\zeta) \right]
\end{equation}
with $ K(\zeta) = e^{-|\zeta|}\,$. Comparing it with the expression (\ref{A1}), we have that
$$ \mathcal{L}\left(T_\tau q, t+\tau \right) = \frac12\,\dot q^2(\tau) -\frac{\omega^2}2\, q^2(\tau) + \frac{g} 4 \, q(\tau)\,\int_\RR \D \zeta \,K(\zeta)\, q(\tau-\zeta) $$
Hence the definition (\ref{L2z}) yields
\begin{equation}  \label{OS2}
\Lambda(q,t,\sigma,\zeta) = \dot q(\sigma)\,\dot\delta(\sigma-\zeta) +\left[-\omega^2\,q(\sigma) +\frac{g}4\,(K\ast q)_{(\sigma)}\right]\,\delta(\sigma-\zeta)+ \frac{g}4\,q(\sigma)\,K(\sigma-\zeta)  \,,
\end{equation}
where $\,K\ast q\,$ is the convolution, and the Lagrange equations are
\begin{equation}  \label{OS3}
\Psi(q,t,\zeta) := \int_\RR \Lambda(q,t,\sigma,\zeta) \,\D\sigma \equiv -\ddot q(\zeta) - \omega^2 q(\zeta) +\frac{g}2\,(K\ast q)_{(\zeta)}
\end{equation}
which are non-local due to the convolution product.

Substituting (\ref{OS2}) in the definition (\ref{L8a}), after a bit of algebra, we obtain the canonical momentum
\begin{equation}  \label{OS4}
P(q,t)_{(\sigma)} = \delta(\sigma)\,\dot q(\sigma) + \frac{g}4\,\theta(\sigma)\,(K\ast q)_{(\sigma)} - \frac{g}4\,\int_0^\infty \D\zeta\,K(\zeta-\sigma)\,q(\zeta)
\end{equation}

We must now put in parametric form the dynamic submanifold $\mathcal{D}^\prime$ given by the constraints (\ref{OS3}):
\begin{equation}  \label{OS5}
 \ddot q(\zeta) + \omega^2 q(\zeta) -\frac{g}2\,(K\ast q)_{(\zeta)} = 0 \,,
\end{equation}
where $K(\zeta)=e^{-|\zeta|}\,$. On differentiating twice with respect to $\zeta$ and including that 
$$ \frac{\D^2 e^{-|\zeta|}}{\D\zeta^2} = e^{-|\zeta|}- 2 \,\delta(\zeta) \,,$$
we have that the constraints imply that
\begin{equation}  \label{OS6}
q^{(iv)} + (\omega^2-1)\,\ddot q +(g -\omega^2) q =0
\end{equation}
That is, any solution of the non-local equation (\ref{OS5}) is also a solution of the differential equation (local) (\ref{OS6}), but the converse is not necessarily true.  
The general solution of (\ref{OS6}) starting at $(q,t)$ is 
\begin{equation}  \label{OS7}
(q,t) \longrightarrow \left(T_\zeta q, t+\zeta\right) \,, \qquad {\rm with} \qquad 
q(\zeta) = \sum_{j=1}^4 A^j \,e^{\zeta \,r_j} \,,
\end{equation}
where $\,r_j \,$ are the roots of the characteristic equation  $\;  r^4 +(\omega^2 - 1) r^2 + g-\omega^2 = 0 \,$; that is 
\begin{equation}  \label{OS8}
r_{\pm\pm} = \pm\, r_\pm\,, \qquad \quad r_\pm = \sqrt{\frac{1-\omega^2}2 \pm \sqrt{\Delta}} \,,\qquad {\rm with} \qquad \Delta = \frac{(\omega^2+1)^2}4 - g 
\end{equation}

For such a function $q(t)$ to be a solution of (\ref{OS5}), the convolution $K\ast q\,$  must exist, which implies that $ \; \int_{-\infty}^\infty \D\tau\, e^{-|\tau| + \tau r_j} < +\infty\,$, that is $\,|\realp (r_j) | < 1\,$. The general solution of (\ref{OS5}) is then
\begin{equation}  \label{OS9}
q(\sigma) = \sum_{j} A^j \,e^{\sigma\, r_j} \,, \quad \mbox{for those $r_j$ such that} \quad |\realp (r_j) | < 1
\end{equation}
This is the parametric equation of the dynamic submanifold $\mathcal{D}^\prime$ and the coordinates are $(A^j,t)\,$. 

Then substituting (\ref{OS9}) in (\ref{OS4}), we obtain the canonical momentum   
\begin{equation}  \label{OS10}
P(q,t)_{(\sigma)} = \sum_j A^j \,\left(r_j \,\delta(\sigma) + \frac{g}4\,\frac{e^{-|\sigma|}}{r_j + \sign(\sigma)} \right)
\end{equation}

\begin{figure}[htb]
\begin{center}
\includegraphics[width=10cm]{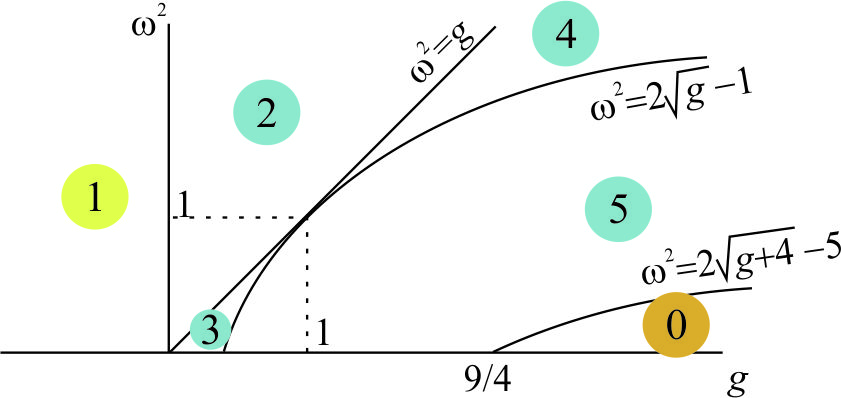}
\end{center}
\caption{Regions in the parameter space $(g,\omega^2)$}  \label{F1}
\end{figure}

In Figure \ref{F1} the parameter space is divided into several regions according to the number and type of roots $r_j$, which we list in the following table:

\begin{center}
\renewcommand{\arraystretch}{1.5}
\begin{tabular}{|c|c||c|c||c|}
\hline
\multicolumn{2}{|c|}{Region} & real $r_j$ & imaginary $r_j$ & $|\realp (r_j) | < 1$ \\
\hline
0 & $\omega^2 \leq 2\,\sqrt{g+4} - 5 $ & 0 & 0 & 0 \\
\hline
1 & $g \leq 0$ & 2 & 2 & 2 \\
\hline
2 & $\omega^2 \geq g > 0$ & 2 & 2 & 4 \\
\hline
3 & $ 2 \sqrt{g}-1 \leq \omega^2 < \min\{g,\,1\}$ & 4 & 0 & 4 \\
\hline
4 & $ \max\{1,\,2 \sqrt{g}-1\} < \omega^2 < g $ & 0 & 4 & 4\\
\hline
5 & $ 2\sqrt{g+4}-5 < \omega^2 <  2 \sqrt{g}-1$ & 0 & 0 & 4 \\
\hline
\end{tabular}
\end{center}

\paragraph{Stability.} Discussing the stability of the solutions of (\ref{OS6}) is a simple task because it is a linear equation with constant coefficients. The solutions are stable if the characteristic roots are simple and their real part is non-positive \cite{Pontriaguine}, \cite{Boyce}. 
As the characteristic equation only contains even powers of $r$, its roots come in pairs with different signs. Hence the solutions are stable if all characteristic roots are simple and imaginary, namely, whenever the parameters $(g,\omega^2)$ lie in regions $1$ or $4$ in Figure \ref{F1}\footnote{In region $1$, we have two real characteristic roots with $|\realp (r_j) | > 1$, therefore they do not contribute to the solution of (\ref{OS5}). They do not affect stability.}.

The latter is in open contradiction with the widespread belief that non-local Lagrangians, and local higher-order Lagrangians as well, suffer from the Ostrogradskian instability \cite{Woodard2015}: as energy (the Hamiltonian) is not bounded from below, the system is unstable. In our view, this inference results from flawed reasoning that consists of taking a sufficient condition of stability, namely ``energy is bounded from below'', as a necessary condition. We shall illustrate it later in this Section.

%%%%%%%%%%%%%%%%%%%%%%%%%%%%%%%%%%%%%%%%%%%%%%%%%%%%%%%
\subsubsection{The reduced Hamiltonian formalism when ${\rm dim}\,\mathcal{D}^\prime=5$\label{S5.2.1} }
Substituting (\ref{OS2}) in (\ref{L13}), we obtain the contact form  
$\omega^\prime \in \mathcal{D}^\prime$
$$ \omega^\prime= \omega - \delta h \wedge \delta t\,, \qquad {\rm where} \qquad 
\omega = \sum_{j,k} \left(r_j + \frac{g(r_k-r_j)}{2 (r_k^2-1)(r_j^2-1)} \right)
\,\delta A^j \wedge \delta A^k  $$
is the (pre)symplectic form\footnote{The question of whether it is symplectic (non-degenerate) or merely presymplectic is addressed later.} which can be easily factored as
\begin{equation}  \label{OS10a}
\omega = \left(\sum_j r_j \,\delta A^j\right) \wedge \left(\sum_k \,\delta A^k \right) + g\,
\left(\sum_j \frac{\delta A^j}{1-r_j^2}\right) \wedge \left(\sum_k \,\frac{ r_k \delta A^k}{1-r_k^2} \right)\,.
\end{equation}
Similarly, the reduced Hamiltonian is derived by combining (\ref{OS10}), (\ref{L12}), and (\ref{OS2}), and it is
\begin{equation}  \label{OS10b}
h = \frac12 \dot q^2 + \frac{\omega^2}2\,q^2 -\frac{g}2 \,\sum_{j,k} A^j A^k\,\frac{1+r_j r_k - r_j^2-r_k^2}{(1- r_j^2)(1-r_k^2)}
\end{equation}

The coordinates $A^j$ are related to the initial data $\; q=q(0)\,, \quad  \dot q=\dot q(0) \ldots \quad  \tridq=\tridq(0)\;$ through equation (\ref{OS7}) which implies that
$$ \sum_j r_j^n \,A^j = q^{(n)} \,, \qquad \quad n= 0 \ldots 3 \,.$$
Solving this for $A^j$ and substituting the result in (\ref{OS10a}), we obtain that the (pre)symplectic form is
$$\omega = \delta\dot q \wedge \delta q + \frac1g\,\delta(\omega^2\, q +\ddot q) \wedge \delta (\omega^2\,\dot q + \tridq) \,, $$
which is obviously non-degenerate and therefore symplectic. Moreover, two pairs of canonical coordinates are apparent, namely  
\begin{equation}  \label{OS11}
 q  \,, \qquad  \quad p = \dot q \,; \qquad \quad \pi = \frac1{\sqrt{g}}\,(\omega^2\, q +\ddot q)  \,; \qquad \quad \xi = \frac1{\sqrt{g}}\,(\omega^2\,\dot q + \tridq) \,,
\end{equation}
in terms of which the symplectic form is
\begin{equation}  \label{OS12}
 \omega = \delta p \wedge \delta q + \delta \pi \wedge \delta \xi 
\end{equation}
It has an associated Poisson bracket \cite{Marsden}, and the non-vanishing elementary PBs are
$$\{q,p\}=\{\xi,\pi\} = 1   $$

Similarly we obtain that the reduced Hamiltonian (\ref{OS10b}) is 
\begin{equation}  \label{OS13}
 h = \frac12\, p^2 + \frac{\omega^2}2\,q^2 - \sqrt{g}\,q\,\pi + \frac12\,\pi^2 - \frac12\,\xi^2
\end{equation}
and it can be easily checked that the Hamilton equations are equivalent to the fourth-order equation (\ref{OS6}). 

It can also be checked that the change of coordinates
$$ q= \tilde{q}_0\,, \qquad p = \tilde{p}_0 + \frac{g}2\,\tilde{q}_1 \,, \qquad \sqrt{g}\,\pi = \tilde{p}_1 + \frac{g}2\,\tilde{q}_0
\,, \qquad \xi = \tilde{q}_1\,\sqrt{g} $$ 
transforms the Hamiltonian (\ref{OS12}) and the symplectic form (\ref{OS12}) into the Hamiltonian system derived in \cite{Kamimura2004} for the same non-local oscillator. 

\paragraph{Stability.} As commented before, if the parameters $(\omega^2,g)$ lie in subregion $4\,$ in Figure \ref{F1}, then the system is stable: a small perturbation in the initial data causes that the perturbed solution oscillates slightly around the non-perturbed solution; the oscillation does not blow up although it does not fade away asymptotically either.

However, the energy, which is an integral of motion, is not bounded from below. 
As commented before, this does not imply that the system is unstable. Indeed, the property that ``energy is bounded from below'' is a sufficient, but not necessary, condition for stability.

\subsubsection{The reduced Hamiltonian formalism when ${\rm dim}\,\mathcal{D}^\prime=3$\label{S5.2.2} } 
When the parameters fall in region 1 in Figure \ref{F1}, ${\rm dim}\,\mathcal{D}^\prime=2+1$. Only two characteristic roots $\pm\,r_-\,$ contribute to the solution (\ref{OS9}) of equation (\ref{OS6}). By a similar procedure as in the previous Section \ref{S5.2.1}, we can derive the contact form
\begin{equation}  \label{OS14}
 \omega^\prime = \delta p \wedge \delta q - \delta h \wedge \delta t \,, \qquad {\rm with} \qquad p = M\,\dot q 
\end{equation}
and the Hamiltonian
\begin{equation}  \label{OS15}
  h = \frac{p^2}{2 M} + \frac{K}2\, q^2\, \qquad {\rm with} \qquad M = 1 - \frac{g}{x_-^2}  \qquad {\rm and} \qquad K = \omega^2 - g\, \frac{2x_- - 1}{x_-^2}  \,, 
\end{equation}
that is an oscillator with frequency 
\begin{equation}  \label{OS16}
 \tilde\omega^2 = \frac{\omega^2 x_-^2 - g (2 x_- - 1)}{x_-^2 - g} \,, \qquad \qquad x_- = \frac{1+\omega^2}2 + \sqrt{\frac{(1+\omega^2)^2}4 - g} 
\end{equation}
This  corresponds to the  Hamiltonian (43-45) in \cite{Kamimura2004} for the perturbative sector.

%%%%%%%%%%%%%%%%%%%%%%%%%%%%%%%%%%%%%%%%%%%%%%%%
%\input{Retarded-NonLocalOscillator-v4.tex}
\subsection{``Delayed'' non-local harmonic oscillator \label{S5.3}}
Consider the action  
\begin{equation}  \label{RO1}
S = \int_{-\infty}^\infty \D t\,\left[\frac12\,\dot q^2(t) -\frac12\,q^2(t) + \kappa\,q(t+T)\,q(t)\right]  
\end{equation}
The non-local Lagrangian is \cite{Woodard2000} $\;\displaystyle{
\mathcal{L}\left(T_\tau q,\tau\right) = \frac12\,\dot q^2(\tau) -\frac12\,q^2(\tau) + \kappa\,q(\tau+T)\,q(\tau)} \;$, and, applying (\ref{L2o}-\ref{L2z}), we have that
\begin{equation}  \label{RO2}
\Lambda(q,t,\tau,\sigma) = - \dot q(\tau)\,\dot\delta(\sigma-\tau) -  q(\tau)\,\delta(\sigma-\tau) +  \kappa\,q(\tau+T)\,\delta(\sigma -\tau) +  \kappa\,q(\tau)\,\delta(\sigma- \tau-T) 
\end{equation}
The Lagrange equation (\ref{L2z}) is 
\begin{equation}  \label{RO3}
\Psi(q, t,\sigma) \equiv -\ddot q(\sigma) - q(\sigma)+ \kappa\,q(\sigma+T) + \kappa\,q(\sigma-T) = 0
\end{equation}
that is, acceleration is proportional to the instantaneous displacement from the origin and also to the displacement $q(t+T)$ at an ``advanced'' time and the ``retarded'' displacement $q(t-T)$. The advanced-retarded symmetry is a consequence of the Lagrangian character of the system.

Then, substituting (\ref{RO2}) into (\ref{L13}), we obtain the contact form
\begin{equation}  \label{RO5}
\omega^\prime = - \delta h  \wedge \delta t + \omega \,, \qquad{\rm where}\qquad  \omega := \delta\dot q(0) \wedge \delta q(0) + \kappa\,\int_0^T \D\sigma\, \delta q(\sigma-T) \wedge \delta q(\sigma)
\end{equation}
is the (pre)symplectic\footnote{At this stage, we cannot assure that $\omega$ is non-degenerate and then symplectic. To ascertain this point, it requires further analysis of the dynamical space.} form and, proceeding similarly with (\ref{L12}), we also obtain the Hamiltonian  
\begin{equation}  \label{RO6} 
h(q,t) = \frac12\,\dot q^2(0) + \frac12\,q^2(0)- \kappa\,q(0)\,q(T) + \kappa\,
\int_0^T \D\sigma\,q(\sigma-T)\,\dot q(\sigma) 
\end{equation}

\subsubsection{The symplectic form \label{S5.3.1}} 
So far, we have not used the Lagrange constraint (\ref{RO3}) yet. As it is a linear equation, the general solution can be expressed as a superposition of exponential solutions like $\;\displaystyle{q(\tau) = e^{i \,\alpha\tau}}\,$, which, substituted in (\ref{RO3}), implies that the factor $\alpha$ in the exponent may take complex values and fulfills the spectral equation
\begin{equation}  \label{R2O7}
1 - \alpha^2 - 2\,\kappa\,\cos(\alpha T)= 0  \,.
\end{equation}
The general solution is thus
\begin{equation}  \label{R2O8}
q(\sigma) = \sum_{\alpha\in\mathcal{S}} A_\alpha\,e^{i\,\alpha\sigma} \,,
\end{equation}
where $\mathcal{S}$ denotes the set of all complex solutions. It is a countable infinite set and, as it can be easily checked, if $\alpha$ is a solution, then $\pm \alpha$ and their complex conjugates $\pm \overline{\alpha}$ are solutions as well. Because $q(\sigma)$ is a real function, the complex spectral coefficients are constrained to
$$ A_{-\overline{\alpha}} = \overline{A_\alpha} \,.$$
Furthermore, the series on the right-hand side of (\ref{R2O8}) should be summable. For non-real $\alpha = x + i\,y\,$, the spectral component $e^{i\,\alpha\sigma} $ blows up at one of the ends, $\sigma\rightarrow \pm\infty$, but this does not affect the local summability. Indeed, it can be proved that, for large values of the real and imaginary parts of $\alpha$,
$$ |x_n| \sim \frac{2 \pi}T\,n  \,, \qquad {\rm and} \qquad |y_n| \sim \frac2T\,\log n \,,\qquad n\in\NN \,.$$
As a consequence, for any $K>0$, it exists $N(K)$ such that $-|x_n| +|y_n|K \leq 0$, for all $n \geq N(K)\,$, and therefore  
$$ \left|  A_{\alpha}\,e^{i\,\alpha\sigma}  \right| = \left|  A_{\alpha}\,e^{-y\sigma}  \right| \leq \left|  A_{\alpha}\,e^{|x|}  \right|\cdot e^{-|x| +|y_n| K} \leq \left|A_{\alpha}\right| \,e^{|\realp \alpha|}  \,, \qquad - K \leq \sigma\leq K $$
Then, provided that $\sum \left|A_{\alpha}\right|\,e^{|\realp \alpha|} < \infty\,$, the series (\ref{R2O8}) converges because it has a convergent majorant.

To sum up, the dynamic space $\mathcal{D}$ consists of all $\,q(\sigma)\,$ of the form (\ref{R2O8}) such that $A = \{A_{\alpha}\,|\; \alpha\in\mathcal{S}\;\;{\rm and}\;\;\sum \left|A_{\alpha}\right|\,e^{|\realp \alpha|} < \infty \} \,$. This suggests that $\mathcal{D}$ is a Banach space with the norm $\;\|A\|:= \sum \left|A_{\alpha}\right|\,e^{|\realp \alpha|}\,$, which may help address a further formalization of the method.

Substituting (\ref{R2O8}) in  (\ref{RO5}), we have that 
$$ \omega = \sum_{\alpha,\,\beta \in\mathcal{S}} \omega_{[\alpha\beta]} \,\delta A_\alpha \wedge \delta A_\beta \,,$$
where
\begin{eqnarray*}
{\rm if}\;\; \alpha+\beta\neq 0\,, & \quad & \displaystyle{\omega_{\alpha\beta} = i\,\alpha - i\,\kappa\,\frac{e^{i \beta T}- e^{- i \alpha T}}{\alpha + \beta} } \\[2ex]
{\rm if}\;\; \beta=-\alpha \,, & \quad & \displaystyle{\omega_{\alpha\,-\alpha}  =i\,\alpha + \kappa T\,e^{- i \alpha T} } 
\end{eqnarray*}
After a short calculation that includes antisymmetrization and equation (\ref{R2O7}), we obtain that the only non-vanishing coefficients are $\omega_{[\alpha\,-\alpha]}$, and we arrive at
\begin{equation}  \label{R2O9}
\omega = \sum_{\alpha\in\mathcal{S}^+} 2\,i\,\left(\alpha - \kappa T\,\sin(\alpha T)\right) \,\delta A_\alpha \wedge \delta A_\beta
\end{equation}
where $\mathcal{S}^+ = \{\alpha\in \mathcal{S}\,|\;\realp(\alpha)> 0,\;\; {\rm or}\;\; \realp(\alpha)= 0,\;\; {\rm and}\;\;\imap(\alpha)>0\}$. 

It is evident from the latter expression that $\omega$ is non-degenerate, hence  symplectic, and also that $A_\alpha\,, \; A_{-\alpha}$, $\;\, \alpha \in\mathcal{S}^+$, are proportional to a pair of canonical coordinates. Moreover, as the matrix of Poisson brackets between pairs of coordinates is the inverse matrix of the coefficients of the symplectic form in those coordinates, we have that the non-vanishing elementary Poisson brackets are:
\begin{equation}  \label{R2O10}
\left\{ A_{\alpha}, A_{-\alpha} \right\} = \frac{i}{2 \left(\alpha - \kappa T\,\sin(\alpha T)\right)} 
\end{equation}
  
To have the Hamiltonian, we must substitute (\ref{R2O8}) in (\ref{RO6}), and we obtain 
$$ h = \sum_{\alpha,\,\beta \in\mathcal{S}} h_{(\alpha\beta)} \, A_\alpha \,A_\beta \,,$$
where 
\begin{eqnarray*}
{\rm if}\;\; \alpha+\beta\neq 0\,, & \quad & \displaystyle{h_{\alpha\beta} = \frac{1-\alpha\beta}2 -\kappa\,e^{i \alpha T} + \kappa\, \beta\,\frac{e^{i \beta T}- e^{- i \alpha T}}{\alpha + \beta} } \\[2ex]
{\rm if}\;\; \beta=-\alpha \,, & \quad & \displaystyle{h_{\alpha\,-\alpha}  =\frac{1+\alpha^2}2 -\kappa\,e^{i \alpha T} - i\,\kappa\,\alpha T\,e^{-i \alpha T}}
\end{eqnarray*}
Similarly as above, we find that $\,h_{(\alpha\beta)} \neq 0\,$ only if $\beta= -\alpha\,$ and also that
$$ h_{(\alpha\,-\alpha)} = 2\,\alpha\,\left(\alpha - \kappa T\,\sin(\alpha T)\right) \,, $$
Finally, the Hamiltonian can be written as
\begin{equation}  \label{R2O11}
h =  \sum_{\alpha, \in\mathcal{S}^+} 2\,\alpha\,\left(\alpha - \kappa T\,\sin(\alpha T)\right) \,A_\alpha \,A_{-\alpha} 
\end{equation}
and the Hamilton equations that follow from the latter with the Poisson brackets (\ref{R2O10}) are
$$ \mathbf{h} A_\alpha = \{A_\alpha, h\} = \frac{\partial h}{\partial A_{-\alpha}}\,\{A_\alpha, A_{-\alpha}\} = i\,\alpha \,A_\alpha \,,$$
as expected ($\mathbf{h}$ is the generator of the Hamiltonian flow).

\subsection{$p$-adic particle  \label{S5.4}}
We now consider the Lagrangian  $\;\displaystyle{ L = -\frac12\,q\,e^{-r \partial_t^2}\,q + \frac1{p+1}\,q^{p+1}  }\,$, which results from neglecting the space dependence in the Lagrangian
$$  L_s = -\frac12\,\phi\,e^{r \Box}\,\phi + \frac1{p+1}\,\phi^{p+1}\,\delta(t-t^\prime) \,, \qquad r=\frac12\,\log p $$
that describes the $p$-adic open string \cite{Sen2000}, \cite{Moeller2002}, \cite{Kamimura2004} ($p$ is a prime integer). 

The linear operator $\displaystyle{\,e^{-\partial_t^2}\,q}$ can be treated as an infinite formal Taylor series that includes the coordinate derivatives at any order or, alternatively \cite{Volovich}, \cite{Treves}, as the convolution 
\begin{equation}  \label{pS1}
e^{-\partial_t^2}\,q(t) \equiv G\ast q (t) = \int_{\RR} \D t^\prime\,G(t^\prime)\,q(t-t^\prime) \,,\qquad {\rm where} \qquad 
G(t) = \frac1{2\sqrt{\pi r}}\,e^{- \frac{t^2}{4 r}}  \,, 
\end{equation}
Using this, the action integral for the above Lagrangian can be written as
\begin{equation}  \label{pS2}
S = \int_{-\infty}^\infty \D t\,\int_{-\infty}^\infty \D t^\prime  \left[-\frac12\,G(t-t^\prime)\,q(t)\,q(t^\prime) + \frac1{p+1}\,q^{p+1}(t)\,\delta(t-t^\prime)  \right] \,,
\end{equation}
which has the form (\ref{A1}) provided that we take
$$ \mathcal{L}\left(T_\tau q,\tau\right) := -\frac12\, q(\tau)\,(G\ast q)_{(\tau)} + \frac1{p+1}\,q^{p+1}(\tau) \,.$$
For this Lagrangian, the functional derivative in equation (\ref{L2o}) is 
\begin{equation}  \label{pS3}
\lambda(q,\tau,\sigma) = \delta(\tau-\sigma)\,\left[-\frac12\,(G\ast q)_{(\tau)} +q^p(\tau) \right] - \frac12\,q(\tau)\,G(\tau-\sigma)
\end{equation}
and, as the Lagrangian does not explicitly depend on $t$, the functional derivative $\Lambda(q,t,\tau,\sigma)\,$ ---equation (\ref{L2z})--- is  $\;
 \Lambda(q,t,\tau,\sigma) = \lambda(q,\tau,\sigma) \,$.

The Lagrange equation (\ref{L2z}) easily follows and has the form of the convolution equation
\begin{equation}  \label{pS4}
G \ast q = q^p  
\end{equation}
According to Vladimirov and Volovich \cite{Volovich}, the only solutions in the space of tempered distributions are 
\begin{equation}  \label{pS4a}
 q_0(\tau) = \left\{\begin{array}{ll}
                   \pm 1, \; 0\,, \qquad &\; \mbox{if $p$ odd} \\
									  \hspace*{.7em} 1, \; 0\,, \qquad &\; \mbox{if $p$ even}
										\end{array}   \right.  
\end{equation}										
If we relax the condition of being a tempered distribution, the convolution equation admits other solutions. We shall concentrate on perturbative solutions around $q_0(\tau)$, namely
\begin{equation}  \label{pS4b}
 q(\tau) = q_0(\tau) + \kappa\,y(\tau) \,, 
\end{equation}										
where $\kappa \ll 1$ is the expansion parameter. Substituting this in (\ref{pS4}) and using the Newton formula for $q^p(\tau)$, we arrive at
\begin{equation}  \label{pS5}
G \ast y - p\,q_0^{p-1}\, y = \kappa\, F \,, \qquad {\rm with } \qquad F := \sum_{n=0}^{p-2} {p\choose n+2}\,q_0^{p-n-2}\,\kappa^n\,y^{n+2}
\end{equation}

For the sake of simplicity, we will concentrate on the case $p=2$, which implies that $q_0(t)=1$ and $\,F = y^2\,$ (the general case involves no additional conceptual difficulty). Then, writing $\; \displaystyle{y = \sum_{n=0}^\infty \kappa^n y_n}\;$ and expanding equation (\ref{pS5}) as a power series of $\kappa$ leads to
\begin{equation}  \label{pS6}
G \ast y_n - 2\, y_n = \sum_{m+k=n-1}\,y_k \,y_m\,, \qquad \quad n\geq 0
\end{equation}
which provides an iteration sequence to be solved order by order: at each level the equation to be solved is linear and contains a non-homogeneous term that depends on the lower order solutions. The general solution is the result of adding a particular solution to the general solution of the homogeneous equation. 

At the lowest order $\,n=0\,$, it reads
\begin{equation}  \label{pS6a}
G \ast y_0 - 2\, y_0 =  0  \,,
\end{equation}
and admits exponential solutions, $\;y_0(\tau) = e^{i \alpha \tau}\,$, where the exponent factor $\alpha$ is a root of the spectral equation
\begin{equation}  \label{pS7}
f(\alpha) := e^{-\alpha r^2} - 2 = 0
\end{equation}
The general solution of (\ref{pS6a}) is therefore
\begin{equation}  \label{pS7a}
  y_0(\tau) = \sum_{\alpha\in\mathcal{S}} Y_\alpha \,e^{i \alpha \tau} \,, \qquad {\rm where} \qquad \mathcal{S} = \left\{\alpha\in \CC\,|\; f(\alpha) = 0 \right\}
\end{equation}

Substituting this in the following order, $n=1$, of equation (\ref{pS6}), we obtain that
\begin{equation}  \label{pS6b}
G \ast y_1 - 2\, y_1 =  \sum_{\beta,\gamma\in\mathcal{S}} Y_\gamma Y_\beta \,e^{i (\gamma +\beta) \tau}  \,,
\end{equation}
Its Fourier transform is $\;\displaystyle{ f(\alpha) \,\tilde{y}_1(\alpha) = \sqrt{2\pi}\,\sum_{\beta,\gamma\in\mathcal{S}} Y_\gamma Y_\beta\,\delta(\alpha-\beta -\gamma) }\,, $ hence, inverting the Fourier transform, we arrive at
\begin{equation}  \label{pS7b}
y_1(\tau) = \sum_{\beta,\gamma\in\mathcal{S}} Y_\gamma Y_\beta\,\frac{1}{f(\beta+\gamma)}\,e^{i (\beta +\gamma)\tau} 
\end{equation}
It is worth remarking that the right-hand side is finite because, as it can be easily proved, if $\,\beta,\, \gamma \in \mathcal{S}\,$, then $\,\beta + \gamma \notin \mathcal{S}\,$. Furthermore, it can be easily inferred that $y_n(\tau)$ is a polynomial of degree $n+1$ in the variables $Y_\alpha\,$.

\subsubsection{The symplectic form and the Hamiltonian}
Now we substitute (\ref{pS3}) into (\ref{L13}) to obtain the (pre)symplectic form
\begin{equation}  \label{pS8}
\omega =  \frac12\,\int_{\RR} \D t\,G(t) \,\int_0^t \D\tau\,\delta q(\tau) \wedge \delta q(\tau-t)
\end{equation}
Then, combining (\ref{pS4b}), (\ref{pS7a}), and (\ref{pS7b}), we obtain after a bit of algebra
$$ \delta q(\tau) = \kappa\,\sum_{\alpha} \delta Y_\alpha\,\left(e^{i \alpha \tau} + 2  \kappa \,\sum_{\beta} Y_\beta\, \frac{e^{i(\alpha+\beta) \tau}}{f(\alpha+\beta)} \right) + O(\kappa^3)\,.$$
Using this, the integral in the right-hand side of (\ref{pS8}) yields
\begin{eqnarray}  
& & \int_0^t \D\tau\,\delta q(\tau) \wedge \delta q(\tau-t) = \kappa^2\,\left[\sum_{\alpha}t\,e^{i\alpha t}\,\delta Y_\alpha \wedge \delta Y_{-\alpha} - \sum_{\alpha+\gamma\neq 0} \frac{i \left(e^{i\alpha t} - e^{-i\gamma t}\right)}{\alpha+\gamma}\,\delta Y_\alpha \wedge \delta Y_{\gamma}  \right.
\nonumber\\[1ex] \label{pS9}
 & & \quad  \left.- 2 \kappa\,\sum_{\alpha,\gamma,\beta} \left(\frac{i \left(e^{i(\alpha+\beta) t} - e^{-i\gamma t}\right)}{f(\alpha+\beta)} +\frac{i \left(e^{i\alpha t} - e^{-i(\gamma+\beta) t}\right)}{f(\beta+\gamma)}\right)\,\frac{Y_\beta\,\delta Y_\alpha \wedge \delta Y_{\gamma}}{\alpha+\beta+\gamma}  \right] + O(\kappa^4)\,,
\end{eqnarray}
where we have included that
\begin{equation}  \label{pS9a}
\int_0^t \D\tau\,e^{i \nu \tau - i\mu t} = 
\left\{  \begin{array}{ll}
    \displaystyle{\frac{- i}{\nu}\, \left(e^{i(\nu-\mu) t} - e^{-i\mu t}\right)   } & \quad \nu\neq 0 \\[2ex]
		  t \,e^{-i \mu t} & \quad \nu=0
			\end{array}  \right.  
\end{equation}			
and the fact that, if $\alpha,\,\beta,\,\gamma \in \mathcal{S}\,$, then $\alpha +\beta \notin \mathcal{S}\,$ and $\,\alpha+\beta+\gamma \neq 0\,$.

%%%%%%%%%%%%%%%%%%%%%%%%%%%%%%%%%%%%%%%%%%%%%%%%%%%%%%%%%%%%%%%%%%%%
Then, substituting (\ref{pS9}) in (\ref{pS8}), we can write
$$ \omega = \kappa^2\,\sum_{\alpha} \omega_\alpha\,\delta Y_\alpha \wedge \delta Y_{-\alpha} + \kappa^2\,\sum_{\alpha+\gamma\neq 0} \omega_{[\alpha\gamma]}\,\delta Y_\alpha \wedge \delta Y_{\gamma} + \kappa^3\,\sum_{\alpha,\gamma,\beta} \omega_{[\alpha\gamma]\beta}\,  Y_\beta\,\delta Y_\alpha \wedge \delta Y_{\gamma} +O(\kappa^4) $$
with
\begin{eqnarray*}
 \omega_\alpha &=& \frac12\,\int_\RR \D t \,G(t)\,t\,e^{i\alpha t} = 2\,i\,\alpha \,r \\[1ex]
\omega_{\alpha\gamma}&=& \frac{-i}{2(\alpha+\gamma)}\,\int_\RR \D t \,G(t)\,\left(e^{i\alpha t} - e^{-i\gamma t}\right) = \frac{-i (e^{-r \alpha^2} - e^{-r \gamma^2})}{2(\alpha+\gamma)} = 0 \\[1ex]
\omega_{\alpha\gamma\beta}&=& \frac{-i}{\alpha+\gamma+\beta}\,\int_\RR \D t \,G(t)\,\left(\frac{e^{i(\alpha+\beta)t} - e^{-i\gamma t}}{f(\alpha+\beta)} + \frac{e^{i\alpha t} - e^{-i(\gamma+\beta) t}}{f(\gamma+\beta)} \right) \\[1ex]
 & = & \frac{-i}{\alpha+\gamma+\beta}\,\left(\frac{e^{-r (\alpha+\beta)^2} - e^{-r \gamma^2}}{f(\alpha+\beta)} + \frac{e^{-r \alpha^2} - e^{-r(\gamma+\beta)^2}}{f(\gamma+\beta)} \right) = 0 \,,
\end{eqnarray*}
where the fact that $f(\alpha)=f(\gamma)=f(\beta)=0$ has been included, and we finally arrive at
\begin{equation}  \label{pS10}
\omega = 4\, i \,r\,\kappa^2 \,\sum_{\alpha\in \mathcal{S}^+} \alpha\,\delta Y_\alpha \wedge \delta Y_{-\alpha} + O(\kappa^4)
\end{equation}
where $\;\mathcal{S}^+ = \{\alpha\in \CC\,|\;f(\alpha)=0,\; \;\realp(\alpha)> 0,\;\; {\rm or}\;\; \realp(\alpha)= 0,\;\; {\rm and}\;\;\imap(\alpha)>0\}$.
It is apparent that $\omega$ is non-degenerate, hence symplectic, and that $Y_\alpha, Y_{-\alpha}$ are multiples of a pair of canonical conjugated coordinates. The elementary non-vanishing Poisson brackets are  
\begin{equation}  \label{pS10a}
 \left\{ Y_{\alpha}, Y_{-\alpha} \right\} = \frac{i}{4 \kappa^2 r \alpha} 
\end{equation} 

To obtain the Hamiltonian, we proceed similarly and, substituting  (\ref{pS3}) into (\ref{L12}), we have that
\begin{equation}  \label{pS11} 
h(q) = h^1(q) - \mathcal{L}(q) \,, \qquad {\rm with} \qquad 
h^1(q) = \frac12\,\int_{\RR} \D t\,G(t) \,\int_0^t \D\tau\, q(\tau) \dot{q}(\tau-t) 
\end{equation}
and
\begin{equation}  \label{pS11a} 
\mathcal{L}(q) = \left[\mathcal{L}\left(T_\tau q,t+\tau \right)\right]_{\tau=0} = -\frac12\,q(0)\,(G\ast {q})_{(0)} + \frac13\,q^3(0) 
\end{equation}
Including now equations (\ref{pS4b}), (\ref{pS7a}), and (\ref{pS7b}), the second integral in the right-hand side becomes
\begin{eqnarray*}
 \int_0^t \D\tau\, q(\tau) \dot{q}(\tau-t) & = & \int_0^t \D\tau\,\left[1 + \kappa \sum_{\alpha} Y_\alpha e^{i\alpha \tau}  \right]\cdot\kappa\cdot  \\[1ex] 
 & & \,\left[ \sum_{\gamma} Y_\gamma \,i \,\gamma \,e^{i\gamma(\tau- t)}  + \kappa \sum_{\gamma,\mu} Y_\gamma Y_\mu \,\frac{e^{i(\gamma+\mu) (\tau-t)}}{f(\gamma+\mu)} \,i(\gamma+\mu) \right] + O(\kappa^3) \,,
\end{eqnarray*}
which can be solved using (\ref{pS9a}) and yields
\begin{eqnarray*} 
\int_0^t \D\tau\, q(\tau) \dot{q}(\tau-t) &=& \kappa \sum_{\alpha} Y_\alpha \left(1 -e^{-i\alpha t} \right) + \kappa^2 \,\left[\sum_{\alpha} Y_\alpha Y_{-\alpha}\,\left(-i \alpha\,t\, e^{i\alpha t} \right) + \right. \\[2ex]
  & & \,\left. \sum_{\alpha+\gamma\neq 0} Y_\alpha Y_\gamma \,\left(\frac{\gamma \,e^{i\alpha t}}{\alpha +\gamma} + \frac1{f(\alpha+\gamma)}\right) \,\left(1- e^{-i (\alpha+\gamma) t}\right) \right] +  O(\kappa^3)
\end{eqnarray*}
Substituting the latter into (\ref{pS9}), we can write
$$ h^1 = \kappa\,\sum_{\alpha} h_\alpha\,Y_\alpha + \kappa^2\,\left[\sum_{\alpha+\gamma\neq 0} h_{(\alpha\gamma)}\,Y_\alpha Y_{\gamma} + \sum_{\alpha} k_{\alpha}\,Y_\alpha Y_{-\alpha} \right]  +O(\kappa^3) $$
with
\begin{eqnarray*}
 h_\alpha &=& \frac12\,\int_\RR \D t\,G(t)\,\left(1-e^{-i\alpha t}\right) = \frac12\,\left(1-e^{-r \alpha^2}\right)= -\frac12 \\[1ex]
k_\alpha &=& - \frac12\,\int_\RR \D t\,G(t)\,i\,\alpha\,t\,e^{i\alpha t}= 2\,r\,\alpha^2 \\[1ex]
h_{\alpha\gamma}&=& \int_\RR \D t \,G(t)\,\left( \frac{\gamma \,e^{i\alpha t} }{2 (\alpha+\gamma)} + \frac1{2\,f(\alpha+\gamma)} \right)\cdot
\left(1 - e^{-i(\alpha+\gamma) t}\right) \\[1ex]
 &=& \frac{\gamma \left(e^{-r \alpha^2} - e^{-r \gamma^2}\right)}{2 (\alpha+\gamma)} +  \frac1{2\,f(\alpha+\gamma)}\, \left(1 - e^{-r(\alpha+\gamma)^2}\right) = -\frac{1 + f(\alpha+\gamma)}{2\,f(\alpha+\gamma)} 
\end{eqnarray*}
where the fact that $f(\alpha)=f(\gamma)=0$ has been included, and we finally arrive at 
\begin{equation}  \label{pS12} 
h^1 = -\frac{\kappa}2\,\sum_{\alpha} Y_\alpha + \kappa^2\,\left[\sum_{\alpha} 2\,r\,\alpha^2\,Y_\alpha Y_{-\alpha} - \sum_{\alpha+\gamma\neq 0} \frac{1 + f(\alpha+\gamma)}{2\,f(\alpha+\gamma)} \,Y_\alpha Y_{\gamma} \right] + O(\kappa^3) 
\end{equation}
 
Proceeding similarly with equation (\ref{pS11a}), we arrive at
\begin{equation}  \label{pS12a} 
\mathcal{L}(q)= -\frac16 -\frac{\kappa}2\,\sum_{\alpha} Y_\alpha - \kappa^2\,\sum_{\alpha,\gamma} Y_\alpha Y_{\gamma} \,\frac{1 + f(\alpha+\gamma)}{2 f(\alpha+\gamma)}+O(\kappa^3)
\end{equation}
and the reduced Hamiltonian is
\begin{equation}  \label{pS13} 
h = \frac16 + 4\,r\,\kappa^2\,\sum_{\alpha\in \mathcal{S}^+} \alpha^2\,Y_\alpha Y_{-\alpha} +O(\kappa^3)
\end{equation}
This result agrees with ref. \cite{Kamimura2004} but, based on functional methods, the present derivation is simpler.  

The Hamilton equations that follow from this Hamiltonian with the Poisson brackets (\ref{pS10a}) are
$$ \mathbf{h} Y_\alpha = \{Y_\alpha, h\} = \frac{\partial h}{\partial Y_{-\alpha}}\,\{Y_\alpha, Y_{-\alpha}\} = i\,\alpha \,Y_\alpha \,,$$

%%%%%%%%%%%%%%%%%%%%%%%%%%%%%%%%%%%%
\section{Conclusion}
We have studied dynamical systems ruled by a non-local Lagrangian. Their treatment differs from the local Lagrangians that are usually considered in mechanics textbooks, especially in what concerns the time evolution and the space of initial data. In the local (first-order) case, Lagrange equations form an ordinary differential system and, thanks to the existence and uniqueness theorems: (a) the space of initial data has a finite number of dimensions ---twice as many as degrees of freedom--- (b) the state of the system is given by the instantaneous coordinates and velocities $(q_i,\,\dot q_j)$ and (c) it evolves according to the solution of Lagrange equations for these initial data.

In contrast, the Lagrange equations in the non-local case are of integro-differential type. There is no general theorem of existence and uniqueness of solutions for such a system. Consequently, the picture of a ``state of the system'' that evolves in time according to Lagrange equations breaks down. Each system requires a specific treatment to determine a set of parameters to characterize each dynamic solution. Additionally, this number of parameters may be infinite.

In our approach, we have opted for: (1) taking Lagrange equations as constraints that select the dynamic trajectories, $\mathcal{D}$, as a subclass among all kinematic trajectories, $\mathcal{K}$, and (2) time evolution is the trivial correspondence $\,q(\tau) \rightarrow q(\tau+t)\,$, i. e. a trajectory evolves in time by advancing its initial point an amount $t$ towards the future. Notice that these two facts also hold in the standard local case, but the existence and uniqueness theorem allows to exploit them further. 

We have then proved an extension of Noether's theorem to the case of a non-local Lagrangian, and, inspecting the form of the conserved quantity, we have guessed the definition of the canonical momenta that we have used to set up the Hamiltonian formalism for a non-local Lagrangian. This could not be based on a Legendre transformation in the usual manner because the latter consists of replacing half of the coordinates in the space of initial data, namely the velocities, with the same number of conjugated momenta. In contrast, in the non-local case, the initial data space is infinite-dimensional, and half infinity is infinity. Infinite dimensions cannot be handled with the same tools as finite ones.  

We have started by considering an almost ``trivial'' Hamiltonian formalism on the kinematic phase space $T^\ast\mathcal{K}$. We have then seen that the Hamiltonian vector field is tangent to a submanifold that is diffeomorphic to the dynamic space $\mathcal{D}$, which has permitted us to translate the Hamiltonian formalism in the larger space onto $\mathcal{D}$. Although this translation could also be done following Dirac's method of constrained Hamiltonian system, we have opted for the symplectic formalism because it is more suited and automatic through pullback techniques. 

In this way, we have obtained the formulae for the Hamiltonian and the symplectic form on the dynamic space provided we have suitably parametrized it. This implies studying the Lagrange constraints, which have to be done specifically for each particular case.  

We have then applied our result to some examples that had been studied elsewhere by other methods. Those methods transform the non-local Lagrangian into an infinite order Lagrangian by replacing the whole trajectories in the non-local Lagrangian with a formal Taylor series (that includes all the derivatives of the coordinates) and then deal with it as a higher-order Lagrangian with $n=\infty$. The value of those methods cannot be heuristic unless the convergence of the series is proved or the ``convergence'' for $n\rightarrow\infty$ is adequately defined. Moreover, these methods are cumbersome in that they often imply handling infinite series with many subindices, square $\infty\times\infty$ matrices, formal inverses, regularizations, etc. 
In contrast, our way is based on functional methods and, as it involves integrals instead of series, is much lighter to handle.

\section*{Acknowledgment}
Funding for this work was partially provided by the Spanish MINCIU and ERDF (project ref. RTI2018-098117-B-C22).

\section*{Data availability}
The data that support the findings of this study are available within the article.

\end{document}